\documentclass[preprint,12pt]{elsarticle}

\usepackage{amssymb}
\usepackage{amsmath}
\usepackage{float}
\usepackage{graphicx}
\usepackage[symbol]{footmisc}
\usepackage{placeins}
\usepackage{url}
\usepackage{eurosym}

\journal{International Journal of Approximate Reasoning}

\begin{document}

\begin{frontmatter}



\renewcommand{\thefootnote}{\fnsymbol{footnote}}

\title{On Classifying the Effects  of Policy Announcements on Volatility}
 \author[Gallo]{Giampiero M. Gallo}
 \author[Lacava]{Demetrio Lacava}
 \author[Otranto]{Edoardo Otranto\footnote[1]{Corresponding author}}
 \address[Gallo]{CRENoS and New York University in Florence, email: giampiero.gallo@nyu.edu}
 \address[Lacava]{University of Messina, email: dlacava@unime.it}
 \address[Otranto]{CRENoS and University of Messina, email: eotranto@unime.it}


\begin{abstract}
The financial turmoil surrounding the Great Recession called for unprecedented intervention by Central Banks: unconventional policies affected various areas in the economy, including stock
market volatility. In order to evaluate such effects, by including Markov Switching dynamics within a recent Multiplicative Error Model, we propose a model--based classification of the dates of a Central Bank's announcements to distinguish the cases where the announcement implies an increase or a decrease in volatility, or no effect. In detail, we propose two smoothed probability--based classification methods, obtained as a by--product of the model estimation, which provide very similar results to those coming from a classical {\it k--means} clustering procedure. The application on four Eurozone market volatility series shows a successful classification of 144  European Central Bank announcements.

\end{abstract}




\date{}
\end{frontmatter}


\section{Introduction}
\label{sec:intro}
Since the onset of the Great Recession, many Central Banks resorted to unconventional monetary policy in order to mitigate the consequences that the crisis had on the real economy and financial markets as well. All the related measures were introduced by means of monetary policy announcements; recent literature focused on the consequences for the real economy (e.g. \cite{kapetanios2012assessing,pesaran2016counterfactual}) and for the financial markets, in particular their volatility \cite{shogbuyi2017effect, kenourgios2015intraday, steeley2015effects}, all viewing the effect of the announcements as a constant factor in the models. The actual  strength of an announcement, however, turns out to be a consequence of the conditions in which the measure is adopted, its wording, how it constitutes a surprise relative to the consensus, how divergent expectations are, and so on. The impact on financial markets, in particular on their volatility, is a consequence of the immediate adjustment of the asset prices to the new information,  and the formation of new equilibria after the announcement.

Within the class of Multiplicative Error Models (MEMs) by \cite{Engle:2002}, the recent work by \cite{Lacava:Gallo:Otranto:2020} is a first attempt at measuring the unconventional policy effects as an unobservable component of volatility, distinguishing the effects due to policy implementation  (as represented by a balance sheet--based continuous proxy variable) from those on the announcement day (related to a dummy variable). In what follows, we chose to modify their univariate Asymmetric Composite Model (ACM -- cf. also the previous contribution by \cite{Brownlees:Cipollini:Gallo:2012}): here, rather than resorting to a dummy variable, we take the view that effects of an announcement could be measured in terms of changes in volatility level across two alternative nonobservable regimes;  to this end, we specify volatility dynamics as Markov Switching (MS), without revealing the model when the announcement took place. As a by--product, we suggest a classification rule for the underlying announcements, according to whether they have a notable impact on volatility via a change in regime or they induced a permanence in the same regime. Such a classification is possible only thanks to the MS approach and would not be reproduceable for other volatility models (notably, the plain Asymmetric (A)MEM by \cite{Engle:Gallo:2006}, or the HAR by \cite{Corsi:2009}).

In spite of it not being a customary classification method based on distance measures, our approach is still within the large literature concerning model--based clustering (for a recent update on the state of the art, cf. \cite{Maharaj:Durso:Caiado:2019} and convenient reviews of these methods are in \cite{liao2005clustering}).  Interestingly,  next to the \textit{whole} and \textit{subsequence} areas of analysis, \cite{aghabozorgi2015time} mention also the  \textit{time-point analysis} strain of research, where we place our approach, aimed at detecting both expected and unusual patterns through the identification of dynamic changes in time series features. 

By the same token, our proposal is clearly different from the model--based techniques aimed at developing clustering in volatility of financial markets, such as \cite{caiadocrato2007}, \cite{otranto2008csda}, \cite{delucazucc2011}, \cite{Durso:Cappelli:DiLallo:Massari:2013}, where the classification involves the full time series and not individual observations. The same considerations can be made by comparing our approach with \cite{otranto2015financial}, where an ACM model similar to the one proposed in this paper is adopted, but, again, with the aim to classify the similarities between entire financial time series.

Based on the estimation of our model on time series of realized volatility (for four Eurozone stock indices, i.e., CAC40, DAX30, FTSEMIB and IBEX35), we consider 144 announcements and we classify how such announcements had an effect on the level of each market volatility: the groups we get are named \textit{Plank} (a neutral effect), \textit{Squat} (a decrease in volatility) and \textit{Jump} (an increase in volatility). Built as a simple processing of the smoothed MS probabilities of being in either a low or a high volatility regime, our classification techniques are simple to implement and deliver results very similar to a benchmark {\it k--means} clustering approach.

The paper is structured as follows. We detail our MS time-point analysis approach to classification in Section \ref{sec:MS}, within which the model is presented in Subsection \ref{sec:models} and the proposed classification procedures in Subsection \ref{sec:classification}. The empirical application is contained in Section \ref{sec:data and results}, where we discuss data features and the framework of events occurring in our sample period; estimation results are discussed in Subsection \ref{subsec:results} and the corresponding classifications in Subsection \ref{subsec:clustering}. Finally, Section \ref{sec:conclusion} contains some concluding remarks.

\section{A Markov Switching Approach to Classification}
\label{sec:MS}

\subsection{A Policy Analysis--oriented Modelling Approach}
\label{sec:models}
Volatility modelling exploits the availability of high frequency data, favouring a decoupling of measurement and modelling with respect to the more traditional approach based on GARCH models \citep{Engle:1982,Bollerslev:1986}. The so--called Realized Volatility (RV) is recognized to have better properties in measurement than the outcome of the estimated GARCH--based conditional variance of returns  \citep{Andersen:Bollerslev:Diebold:Labys:2001}. As per forecasting, several conditional models are available for RV; one of them is the MEM, proposed by \cite{Engle:2002}, which takes volatility dynamics in terms of the product of two positive time-varying factors, one representing its conditional mean and the other a positive--valued disturbance. Several improvements allow to capture stylized facts and to accommodate specific cases; in particular the \cite{Engle:Gallo:2006} specification introduces the asymmetric and predetermined variable effects.
In what follows, with an eye to capturing policy effects, we extend the MS--AMEM by \cite{Gallo:Otranto:2015} to additively accomodate policy--induced effects (component $\xi _{t,s_t}$) next to the volatility dynamics (the \textit{base} volatility component $\varsigma _{t}$):
\begin{equation}
\begin{array}{l}
 RV_{t}=\mu _{t,s_t}\epsilon_{t}, \, \epsilon _{t}|\mathbf{I}
_{t-1}\sim Gamma(\vartheta_{s_t} ,\frac{1}{\vartheta_{s_t}})\\
\mu _{t,s_t}=\varsigma _{t}+\xi _{t,s_t}\\
\varsigma _{t}=\omega +\alpha RV_{t-1}+\beta \varsigma
_{t-1}+\gamma D_{t-1}RV_{t-1}\\
\xi _{t,s_t}=\varphi_0+\varphi_{1} s_t+ \delta (E\left(x_{t}|\mathbf{I}_{t-1}\right)-\bar{x})+ \psi \xi _{t-1,s_{t-1}}
\end{array} \label{eq:MScsiACM}
\end{equation}
As with any other MEM, the realized volatility at time $t$, $RV_t$, is seen as the product of a conditional (on the past information set $\mathbf{I}_{t-1}$) expectation term $\mu_t$ times a unit mean error term  $\epsilon_t$ following a  Gamma distribution.\footnote{In suggesting the Asymmetric MEM, \cite{Engle:Gallo:2006} justify the adoption of a Gamma distribution in view of its flexibility and the fact that it nests other notable distributions (such as the exponential and the Chi-square). In a more general framework for MEMs, \cite{Cipollini:Engle:Gallo:2013} show that the first order conditions for a Gamma--based likelihood function coincide with the objective function of a semiparametric GMM based approach.}

In our approach, in line with \cite{Lacava:Gallo:Otranto:2020}, the expected conditional volatility is decomposed as the sum of $\varsigma_t$, evolving as a GARCH--like process (with asymmetric effects tied to the negative sign of past returns captured by a dummy variable, $D_{t-1}$), and a policy--specific and regime--dependent term $\xi _{t,s_t}$, which follows an AR(1) model. The driving variable for this dynamics is $x_t$ -- a proxy for unconventional policy measures, entering the model as the deviation of its conditional expectation from a long term mean $\bar x$ -- which accounts for the Central Banks' balance sheet composition. 

{In the original ACM model by \cite{Lacava:Gallo:Otranto:2020}\footnote{Actually, this model is derived from the general framework in \cite{Brownlees:Cipollini:Gallo:2012}, where the mean of the conditional volatility is the sum of two unobservable components: the particular ACM specification proposed by \cite{otranto2015capturing} to model spillover effects in financial markets is adapted to representing the base volatility and the unconventional policy effect respectively.} the last equation is:
\begin{equation}
\begin{array}{c}
\xi _{t}=\delta (E\left(x_{t}|\mathbf{I}_{t-1}\right)-\bar{x})+ \varphi (\Lambda_{t}-\bar{\Lambda})+ \psi \xi _{t-1},
\end{array}\label{eq:ACM}
\end{equation}
where,  the $\Lambda_t$ term is a dummy variable representing the effect of the announcements and is taken as deviation from its long--term mean, also here; the days of the announcements are not random variables, as they are put in the calendar in advance by the ECB.

Contrary to the approach by \cite{Lacava:Gallo:Otranto:2020}, we are not considering the announcements explicitly in the model via dummy variables; what constitutes the novelty in the present econometric context is that we detect their presence when a change in regime is attributable to a market volatility reaction to the announcement. More in detail,} we extend the ACM model by \cite{Lacava:Gallo:Otranto:2020} in making the policy--specific component $\xi_t$ become $\xi _{t,s_t}$, i.e.,  we  consider a dichotomic discrete latent variable $s_t=0,1$, representing the regime at time $t$.
When $s_t=0$, the time series is in a low volatility regime with intercept $\varphi_0$, and increases by the term $\varphi_1\ge 0$ in the high volatility regime ($s_t=1$). In practice, though, as the regime is not observable, such an intercept turns out to be time--varying, as discussed below. The dynamics of the state variable $s_t$ is driven by a first--order Markov chain, that is:
\begin{center}$Pr(s_{t}=j|s_{t-1}=i,s_{t-2}\ldots
)=Pr(s_{t}=j|s_{t-1}=i)=p_{ij}.$ \end{center}
{Building also on the evidence and the discussion presented in \cite{Gallo:Otranto:2015}, we obtain the important result of expanding the flexibility of the model, since the distribution of the error term now follows a mixture of two Gamma densities.}

Positiveness ($\omega>0$, $\alpha, \beta, \gamma \geq 0$) and stationarity ($\alpha+\beta+\frac{\gamma}{2}<1$ and $\left|\psi\right|<1$) conditions established in the case of the ACM are regime--independent and hence are valid for the MS--ACM as well.

{The Markov process assumption implies the  dependence only on the current state and implicitly assumes that the sojourn distribution is geometric. In principle, we can assume a more general dependence  on the duration of the state or on $p$ lagged states, adopting semi--Markov processes and transition distribution models. In the present framework, however, we rely on customary assumption of market efficiency, based on the tenet that  information is instantly and completely incorporated into the current price. This theory is supported by the empirical evidence with daily data and is consistent with the econometric literature of MS models in the GARCH framework (see, for example, \cite{Gray:1996,Dueker:1997,Klaassen:2002}).}

The likelihood function of the MS--ACM is obtained by means of the so called Hamilton filter and smoother, as described in \cite{Hamilton:1994} (Ch.~22), adopting the approximated solution proposed by \cite{Kim:1994} to solve the path dependence issue, a  computational problem due to the dependence of $\mu_t$ on all past values of $s_t$. In fact, at the end of the recursive Hamilton filter, we would have to keep track of all possible paths obtained by all the combinations of regimes from the first time until the last one, with $2^T$ possible different scenarios. To by--pass this problem,  \cite{Kim:1994} proposes to collapse the 4 possible values of $\mu_t$ at time $t$, obtained at the end of each step of the Hamilton filter,  into $2$ values, averaging and weighing them with the corresponding conditional probabilities $Pr[s_{t-1}=i,s_{t}=j|\mathbf{I}_t]$ (obtained in the same Hamilton filter step). {This approximation is the most popular solution, albeit not exact, causing a bias in the likelihood function; in practice, though, the collapsing procedure does not involve substantial errors in the final estimation: as shown by \cite{Gallo:Otranto:2015} through simulation experiments, the approximated likelihood is satisfactory when the number of observations is above 2000, an attainable goal with daily financial time series. 

As a recent contribution, \cite{Augustyniak:Boudreault:Morales:2018} have generalized and improved the collapsing procedure for MS--GARCH models, implementing optimal particle filters to explicit the likelihood; the particle selection step is performed deterministically and not stochastically, favoring the implementation and reducing the computational effort. The authors compare their approximate simple solution with the simulation--based approach of \cite{Augustyniak:2014}, showing a small bias in practical situations. Moreover, this approach avoids the typical problem of the particle filters, which provide likelihood estimates that are discontinuous functions of the parameters. A recent simulated particle filter approach was proposed by \cite{Wee:Chen:Dunsmuir:2020}, as a modification to the classical setup, obtaining estimated likelihood functions amenable to numerical optimization and able to work also in presence of missing data. The comparison of these different estimation procedures in the MS--AMEM framework is interesting, but is beyond the scope of the present work.}

The so--called smoothed probabilities $P[s_t|\mathbf{I}_T]$ provided by the Hamilton smooth\-er are used to make inference on the regime conditional on the full information available, $\mathbf{I}_T$; a rule of thumb consists of assigning the observation at time $t$ to regime 1 if  $\hat{p}_t\equiv P[s_t=1|\mathbf{I}_T]>0.5$, otherwise to regime 0. In our context, the smoothed probabilities are further used to estimate the intercept in (\ref{eq:MScsiACM}), as a weighted average of $\hat\varphi_0$ and $\hat{\varphi}_1$:
\begin{equation}
\hat{\varphi}_t=\hat{\varphi}_0 (1-\hat{p}_t)+(\hat{\varphi}_0+\hat{\varphi}_1) \hat{p}_t
\label{eq:phi_est}
\end{equation}
where a hat indicates the QML estimate of the parameter. The interesting feature of our model is that the intercept of $\xi _{t,s_t}$, the coefficient $\hat{\varphi}_t$, is time--varying, providing the result that, differently from Eq. (\ref{eq:ACM}), a change in regime will induce a change in the level of the series with varying intensity, without feeding the model the dates of the announcements. This is in contrast with a model relying on a dummy variable for announcements which would force the effect on volatility dynamics to be constant.

When we overlap the dates of the announcements to the estimated results,  we can thus monitor a possible change in the volatility level and its amplitude at each date. As a consequence, we propose two different methods to classify the policy announcements based on the changes of $\hat{\varphi}_t$ when $t$ is an announcement day, relative to the day before.

\subsection{Classification of announcements}
\label{sec:classification}
 In formal terms, the $N$ dates of announcements (i.e., when $\Lambda_t=1$) are selected within the overall time series and, for those, the values $\hat{\varphi}_t-\hat{\varphi}_{t-1}$ are calculated. Notice that, from (\ref{eq:phi_est}):
\begin{equation}
\Delta \hat{\varphi}_\tau \equiv \hat{\varphi}_\tau -\hat{\varphi}_{\tau-1}=\hat{\varphi}_1(\hat{p}_\tau-\hat{p}_{\tau-1})\equiv\hat{\varphi}_1\Delta \hat{p}_\tau \quad \forall ~ \tau=t: \Lambda_t=1,
\label{eq:dprob}
\end{equation}
that is, such announcement effects on the volatility level can be evaluated through the variations in the smoothed probabilities directly. Thus, a first form of classification of the $N$ announcement is to apply a clustering algorithm to obtain groups with similar $\Delta \hat{p}_t$.

However, by the approach detailed in Eq.\eqref{eq:dprob}, one can notice that the announcement effect (a large movement in the intercept) will be estimated to be substantial, the larger the movement between the probability of being in regime 1 at time $t-1$ and the corresponding one at time $t$.
An alternative to a clustering--based method can therefore be suggested, exploiting the customary mapping of smoothed probabilities into a regime classification based on the threshold $\hat{p}_t=0.5$.

A first smoothed probability--based classification (dubbed {\it SP-level}) can be obtained directly from  the position of the probabilities $\hat{p}_t$, respectively, $\hat{p}_{t-1}$, relative to the threshold value $0.5$. In this case we suggest 4 groups, with an immediate interpretation, in the following way:
\begin{enumerate}
    \item {\it No effect and low volatility -- (low) Plank}, if $\hat{p}_t\leq 0.5$ and $\hat{p}_{t-1}\leq 0.5$;
    \item {\it No effect and high volatility -- (high) Plank},  if $\hat{p}_t\geq 0.5$ and $\hat{p}_{t-1}\geq 0.5$;
    \item {\it Decrease in volatility -- Squat}, if $\hat{p}_t\leq 0.5$ and $\hat{p}_{t-1}\geq 0.5$;
    \item {\it Increase in volatility -- Jump}, if $\hat{p}_t\geq 0.5$ and $\hat{p}_{t-1}\leq 0.5$.
\end{enumerate}

A second smoothed probability--based classification (called {\it SP--diff}, since it is based on $\Delta \hat{p}_t$) gives three groups as follows:
\begin{enumerate}
    \item {\it No effect -- Plank}, if $-0.5 \leq \Delta \hat{p}_t \leq 0.5$; this group will contain cases with a moderate effect, with  or without regime change (i.e. irrespective of whether the  threshold is crossed), since subsequent $\hat{p}_t$'s are close to one another;
    \item {\it Decrease in volatility -- Squat}, if $\Delta \hat{p}_t < -0.5$; in this case the volatility at time $t$ is attributed to regime 0 whereas the volatility at time $t-1$ to regime 1, with a sharper  change in the value of the probability by more than 50\%;
    \item {\it Increase in volatility -- Jump}, if $\Delta \hat{p}_t>0.5$; in this case the volatility at time $t$ is attributed to regime 1, whereas the volatility at time $t-1$ to regime 0, and the change in the value of the probability is more than 50\%.
\end{enumerate}
The diff--groups here are different from the level--groups, even if they have the same label, since in this second approach we would classify as a Plank a change in regime between $t-1$ and $t$, not accompanied by a relevant change in $\hat p_t$'s. By both classifications, however, Squats and Jumps come when the MS model produces a sharp mapping into regimes, and therefore should give similar results.

If well designed, the three classifications should provide similar results, and evidence will be provided below; both smoothed probability--based classifications have the advantage of being immediately applicable each time an announcement is provided.\footnote{{Notice that our decoding is local, given that the final clustering depends on the smoothed probabilities derived from the Hamilton filter and Kim smoother.}}

To evaluate the degree of uncertainty attributable to these classification methods based on the smoothed probabilities, we propose a new index. We note that the sharpest assignment of the volatility to state 0, respectively, 1, is obtained when the smoothed probability is exactly equal to 0 or 1. In reference to our distinction between {\it Plank} (P), {\it Squat} (S) and {\it Jump} (J), such a case implies that the difference of the contiguous probabilities would  then be equal to 0 in the case P, (-1) in the case S and (+1) in the case J.\footnote{Note that  the {\it SP--level} and {\it SP--diff} coincide, in such a case.} As a consequence, we suggest the following index:
\begin{equation}
U=\frac{2}{N} \sum_{\tau=1}^{N} \left[\left|\Delta^{(P)}\hat{p}_t\right|+\left|\Delta^{(S)}\hat{p}_t+1\right|+\left|\Delta^{(J)}\hat{p}_t-1\right|\right],
\label{eq:U}
\end{equation}
which varies between 0 (sharpest classification) and 1 (case of maximum uncertainty).\footnote{Since  the differences in the group P assume values between -0.5 and 0.5, in group S between -1 and -0.5, in group J between 0.5 and 1, we multiply the average in Eq. (\ref{eq:U}) by 2 to map the result into $[0,1]$.}

\section{An Empirical Application}
\label{sec:data and results}

\begin{figure}[h]
    \centering
    \includegraphics[scale=0.66]{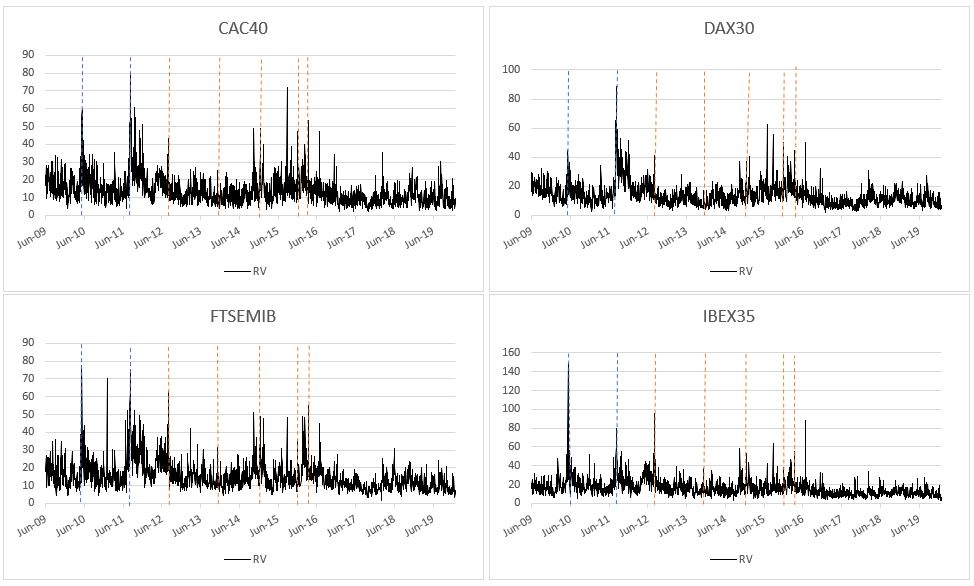}
    \caption{Realized Volatility series and relevant events (red and blue dashed-lines) for financial markets. Sample Period: June 1, 2009 to December 31, 2019}
    \label{fig:rv_series}
\end{figure}

We consider  the annualized realized kernel volatility of four Eurozone stock indices (CAC40, DAX30, FTSEMIB and IBEX35) as provided by the Oxford Man Institute\footnote{Data are available at \url{https://realized.oxford-man.ox.ac.uk/data/download}} for the period from June 1, 2009 to December 31, 2019 (daily data, 2685 observations). Their profiles, shown in Figure \ref{fig:rv_series}, behave similarly between one another, exhibiting bursts of market activity occurring in correspondence with some events of relevance: two remarkable peaks are recognizable in the first part of the sample, coinciding with the flash crash on May 6, 2010 and with the \textit{Black Monday} on August 8, 2011 (depicted as blue--dashed vertical lines in Figure \ref{fig:rv_series}). Similarly, some volatility spikes correspond to monetary policy announcements (red--dashed line); this is the case, for example, of the so--called conventional monetary policy decisions concerning the interest rates (on August 2, 2012; November 7, 2013; December 4, 2014; December 3, 2015) and the \textit{unconventional} policy decision on March 10, 2016, when the Corporate and Public Sector Purchases Programmes (CSPP and PSPP, respectively) were included in the Expanded Asset Purchases Programme (EAPP). In addition, volatility clustering emerges quite clearly throughout, with a long period of low volatility starting in July 2012, whereas short--lived periods of high volatility are observed at the beginning of the sample, corresponding to the Greek sovereign debt crisis in May 2010 and in the mid of 2011 when the sovereign debt crisis exploded spilling over beyond the Eurozone.

We proxy for the implementation effects, the term $E\left(x_{t}|\mathbf{I}_{t-1}\right)$ in Eq.\eqref{eq:MScsiACM}, via the ratio of the amount of securities held for unconventional policy purposes to that employed for conventional policy measures\footnote{Data are obtained from the  ECB website and Datastream.}. The conditional expectation of $x_t$ is estimated through the ARIMA(4,1,1) model, according to a preliminary order identification procedure. Finally, the list of monetary policy announcements consists of $N=144$ events, constructed starting from the ECB press releases\footnote{Available at  \url{https://www.ecb.europa.eu/press/pr/activities/mopo/html/index.en.html}}, each setting  $\Lambda_t=1$.

\FloatBarrier

{In what follows, we chose to set the number of states to 2, after a preliminary investigation among the alternatives. As it is well known, in the MS world competing models cannot be compared with classical tests for the problem of nuisance parameters present only under the alternative hypothesis (\cite{Hansen:1992}). This notwithstanding, we have estimated 3-state MS models for each market and compared them with the 2-state and the no MS cases, relying on the information criteria as a tool for model selection (\cite{Psaradakis:Sola:Spagnolo:Spagnolo:2009}); across  all series, the 2 state case is preferred in terms of BIC. This result is further confirmed by the results of the 3--state case estimate, where the switching coefficient $ \varphi $ of the third state is not significant.\footnote{Results are available among the supplemental material.}}

\subsection{Estimation results}
\label{subsec:results}

Estimation results of our MS--ACM are shown in Table \ref{tab:MSXi-ACM} for the volatilities of each financial index.\footnote{{Initializing ML estimation from 11 sets across a rather wide range of random starting values for each model, we converge to the same coefficient estimates, up to the third digit.  On average, the computational time of the full estimation procedure, using a GAUSS code running on an I5--7th gen 2.5 Ghz processor, is approximately 2 minutes.}} They support our choice to assume the constant of the $\xi_{t,st}$ component as switching: in particular, the constant in the low volatility regime is equal to zero across indices (a regime without monetary policy announcements), whereas it increases remarkably in the high volatility regime. As a consequence, we removed $\varphi_0$ from all models, and, for the same reasons, the AR coefficient $\psi$. The unconventional policy proxy significantly enters the model with a negative sign, as it is expected to reduce the volatility level, with the strongest impact observed for the IBEX35 (-1.09) and the weakest occurring for the DAX30 (-0.44). As per the probability coefficients, the probability of remaining in the low regime is higher than that of the high volatility regime, leading to an average duration of 1 day (calculated as $\frac{1}{1-p_{ii}}, i=0,1$) in the high regime for all markets, while the duration in the low regime ranges between 14 days (FTSEMIB) and 53 days (DAX30). 

Some model diagnostics reinforces the quality of our approach: in Table \ref{tab:diagnostic} we show a few tests on the residuals. The Ljung--Box statistics, calculated for lags 1, 5, 10,  demonstrate that no substantial autocorrelation remains in the residuals, which also show sample mean equal to 1, in line with the hypothesis about the center of the Gamma distribution. Additionally, a Kolmogorov-Smirnov test fails to reject the null hypothesis (even at a 10\% significance level) that the distribution is a mixture of two Gamma densities weighted by the corresponding ergodic probabilities (derived from the estimated Markov Chain). Furthermore, in order to compare the goodness of fit across alternative models, we built an in-sample Model Confidence Set (\cite{Hansen:Lunde:Nason:2011}); the competing models, next to our MS--ACM (Eq. \ref{eq:MScsiACM}), are the classical AMEM (\cite{Engle:Gallo:2006}), the AMEMX (an AMEM with $x_{t}$ added as a regressor), and the  ACM (Eq. \ref{eq:ACM}). The procedure always selects our MS model as the best model and none of the alternatives falls into the best set. Finally, as a further check, we calculated the cross correlations at lag 1 between the residuals across markets; the results show very low residual lagged cross correlations (all below $0.1$ and most below $0.05$), pointing to the fact of no evidence of dynamic interdependence. Evidently, this issue should be addressed within a multivariate analysis, but this goes beyond the scope of this work. 

\begin{figure}[H]
\centering
\includegraphics[scale=1.2]{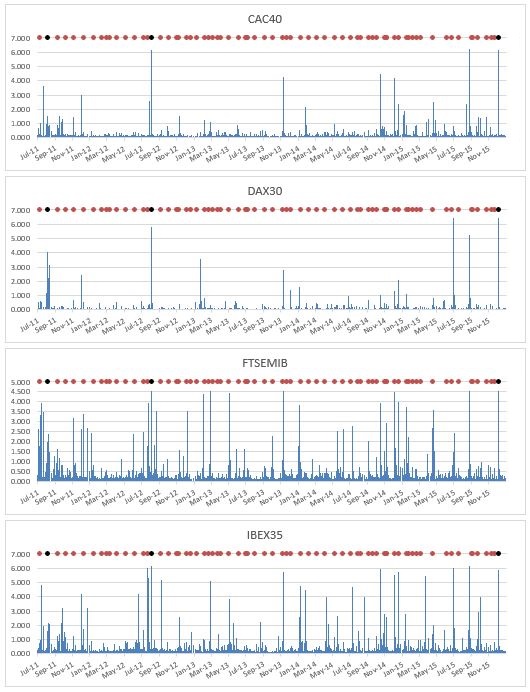}
\caption{$\hat{\varphi}_t$ as a weighted average resulting from the MS--ACM (blue line) and monetary policy announcements (red dots). The period depicted spans July 1, 2011 to December 31, 2015. Black dots correspond to three meaningful announcements (details in the text): August 4, 2011, August 2, 2012, and December 3, 2015.} \label{fig:announ_qx}
\end{figure}

In Figure \ref{fig:announ_qx}, we reproduce the time--varying constant $\hat \varphi_t$, superimposing the dates at which an announcement occurred (an information not provided to the model). In line with an immediate absorption of the news, the series jump in correspondence of monetary policy announcements, more frequently so for the FTSEMIB and the IBEX35, than for the CAC40 and the DAX30. In order to add readability to the results, we consider three meaningful dates, in detail: August 4, 2011, when the ECB gave additional details on a Longer Term Refinancing Operation (LTRO); August 2, 2012, when the ECB communicated to the market that there would have been no changes of interest rates, and December 3, 2015, when, conversely, only the interest rate on deposit facility was decreased by 10bps. Note that these announcements depressed returns in all cases, with the worst loss happening at the unconventional policy announcement (August 4, 2011), between 4.41\% (DAX30) and 6.9\% (FTSEMIB). Interestingly, in the case of the two conventional policy announcements, the estimated probability to be in  the high regime is near 1;  on the day of the unconventional policy announcement (August 4, 2011), the process appears to be in the low regime, an empirical support to the idea that unconventional policies were successful in reducing stock market volatility.\footnote{This fact is also confirmed by the negative sign of the $\delta$ coefficient in Table \ref{tab:MSXi-ACM}.} Finally, we observe a reduction in $\hat{\varphi}_t$ after the announcements, as evidence of the short estimated permanence in the high volatility regime.

{As a general remark, we find support for our motivation to address the very stylized facts of RV within the MEM class: persistence, captured by an ARMA representation of the dynamics; asymmetric response of volatility to the sign of past returns, captured by a significant asymmetric term; lack of serial correlation in the residuals, ascertained by the usual Ljung--Box test statistics; the addition of MS features advanced here allows us to improve the goodness of fit, as discussed before, and a different consideration of quiet and turbulent periods.}

\begin{table}[H]
  \begin{center}
  \begin{footnotesize}
  \caption{Estimation results  (robust sandwich form of the standard errors \cite{White:1982} in parentheses)  of four MS--ACMs relative to different Eurozone volatility series. Sample period: June 1, 2009 - December 31, 2019.}
    \begin{tabular}{l c c c c}
  & CAC40 & DAX30 & FTSEMIB &IBEX35 \\
    \hline
$\omega$ & 0.853   & 0.661   & 1.063   & 1.059   \\
          & (0.107) & (0.072) & (0.122) & (0.079) \\
$\alpha$ & 0.142   & 0.185   & 0.228   & 0.151   \\
          & (0.017) & (0.015) & (0.019) & (0.018) \\
$\beta$  & 0.732   & 0.722   & 0.649   & 0.732   \\
          & (0.024) & (0.019) & (0.024) & (0.019) \\
$\gamma$ & 0.112   & 0.082   & 0.080   & 0.086   \\
          & (0.011) & (0.009) & (0.019) & (0.011) \\
$\delta$ & -0.776   & -0.440   & -0.741   & -1.090   \\
          & (0.131) & (0.104) & (0.157) & (0.121) \\
$\varphi_1$ & 6.273   & 6.420   & 4.552    & 6.152    \\
          & (1.811) & (1.647) & (1.853) & (2.534) \\
$p_{00}$ & 0.964   & 0.981   & 0.928   & 0.943   \\
          & (0.025) & (0.017) & (0.027) & (0.036) \\
$p_{11}$ & 0.222   & 0.303 & 0.337 & 0.313   \\
          & (0.122) & (0.254) & (0.460) & (0.099) \\
$\theta_0$ & 8.852   & 11.248   & 15.034   & 11.992   \\
          & (0.379) & (0.509) & (1.375) & (0.756) \\
$\theta_1$ & 3.271   & 2.595   & 4.112   & 3.777   \\
          & (0.737) & (0.767) & (1.211) & (0.683) \\
          \hline
   \end{tabular}\label{tab:MSXi-ACM}
    \end{footnotesize}
         \end{center}
\end{table}

\begin{table}[h]
  \centering
  \caption{Residual diagnostics of four MS-ACMs relative to different Eurozone volatility series: p-values of Ljung--Box statistics for different lags, mean of residuals (standard deviation in parentheses) and Kolmogorov-Smirnov statistic.}\label{tab:diagnostic}
    \begin{tabular}{lrrrr}
          & \multicolumn{1}{c}{CAC40} & \multicolumn{1}{c}{DAX30} & \multicolumn{1}{c}{FTSEMIB} & \multicolumn{1}{c}{IBEX35} \\ \hline
    Ljung--Box 1 & 0.711 & 0.161 & 0.403 &0.220 \\
    Ljung--Box 5 & 0.263 & 0.121 & 0.449 & 0.087  \\
    Ljung--Box 10 & 0.364 & 0.276 & 0.639 & 0.340 \\ \hline
    mean         & 0.994 & 0.998 & 0.996 & 0.994  \\
    & (0.343)&(0.321)&(0.292)&(0.303)\\ \hline
    Kolmogorov-Smirnov & 0.019 & 0.015 & 0.016 & 0.015 \\ \hline
    \end{tabular}\\
    \vspace{2mm}
  \footnotesize{The critical values of the Kolmogorov--Smirnov statistic are 0.024 at a significance level of 0.10, 0.026 at a significance level of 0.05, 0.031 at a significance level of 0.01. }
\end{table}

\FloatBarrier

\subsection{Classification Results}
\label{subsec:clustering}
The estimated smoothed probabilities are used to classify the 144 dates of announcements into the three categories: {\it Plank}, {\it Squat}, and  {\it Jump}. We calculated the variable $\Delta \hat{p}_t$ (see Eq.\eqref{eq:phi_est}) and apply a classical {\it k--means} algorithm, minimizing the sum of squares from points to the assigned cluster centers ({the average of $\Delta \hat{p}_t$ within each group}),  to form three groups. In the first column of Table \ref{tab:announ_regimechanges} we show the number of dates belonging to each group for each market. Most of the announcements do not cause a change in volatility ({\it Plank}); however, a remarkable percentage of the announcements cause a switching from the low to the high volatility for FTSEMIB and IBEX35, that is 15.3\% and 8.3\%, respectively, whereas it is less than 5\% for CAC40 and DAX30. Some important announcements belonging to this category refer both to conventional policies (for example, the decrease of interest rates established on December 3, 2015, when they became negative) and unconventional policies (e.g. the announcement on March 10, 2016, when the amount of securities purchased within the implementation of the EAPP passes from \EUR{60} to \EUR{80} billion per month; the announcement on September 12, 2019, when the ECB decided to run the EAPP as long as necessary).

Finally, a smaller percentage of the announcements caused a switching from the high to the low volatility regimes: some examples are represented by the details on the Covered Bond Purchases Program (CBPP) released on June 4, 2009 (for the CAC40), the Security Market Program (SMP) on May 10, 2010 (for the FTSEMIB) and the announcement on June 8, 2017 concerning details on the EAPP (for the IBEX35).

Our analysis confirms the results in \cite{Lacava:Gallo:Otranto:2020} where the dummy representing the announcements has a positive sign and is more significant for the FTSEMIB and the IBEX35. The announcement effect seems to be more pronounced for the volatility of the Italian and Spanish markets than for the French and German ones, in line with the more stable performance of the latter during this turbulent period.

In the last two columns of Table \ref{tab:announ_regimechanges} we show the clustering obtained with the two smoothed probability--based approaches (we merged the two  {\it Plank} cases in the {\it SP--level} classification, in order to make the results comparable with the other classifications). The outcome seems quite similar across methods, with a larger number of cases identified as \textit{Plank} announcements relative to what the {\it k--means} clustering delivers. Also, considering the {\it SP--level} approach, we can notice that almost all \textit{Plank} cases are identified when  the regime was one of low volatility at time $t-1$.

The centers of each group are around 0 for the {\it Plank} group across classification methods. Some differences are present in the centers of the other two categories, in particular between the {\it k--means} method and the two smoothed probability--based approaches; FTSEMIB seems to be the one market with less sharp classifications when derived from the three methods.

{ In Table \ref{tab:Uindex} we report the values of $U$ for each classification method and each series. In general, {\it SP-diff} presents the lowest $U$ values, but also {\it SP-level} has a very similar performance. }

A formal evaluation of the differences in the classifications obtained with the three alternative methods can be conducted by means of the adjusted Rand index \citep{ran71,ha85}. Such a method is generally used to compare the groups obtained by a certain algorithm with respect to a benchmark clustering; in our case, we use it to verify the similarity of clustering methods by taking possible pairs in turn. The Rand index ranges in [0, 1], and takes on value 1 when the two methods provide the same clustering, and value 0 in the case of maximum difference between them. In Table \ref{tab:rand_methods} we show the values of this index across volatility series; they are always larger than 0.9 (with the slight exception of the comparison between {\it k--means} and {\it SP--diff} for the FTSEMIB, which is 0.85), with value 1 in the case of the two smoothed probability--based methods for the CAC40 and the DAX30. Given that the three alternative methods provide very similar results, the smoothed probability--based solutions receive a good support as an immediate by--product of the model estimation, not requiring further statistical clustering algorithms.

Furthermore, to verify if the announcements are classified in a similar way, this time across the four volatility series, we calculate the Rand index for the same clustering method, but for different markets (Table \ref{tab:rand_countries}); once again, we get a strong agreement in the classification results, with Rand indices always larger than 0.74, with a clear similarity between the CAC40 and the DAX30. Moreover, it seems that the {\it SP--diff} method provides more similar patterns among markets relative to the other two methods. By comparison, the {\it k--means} provides lower values, but still high in terms of similarity.

We can conclude that the smoothed probability--based classifications confirm their good performance in providing reliable results, also in view of their very practical derivation, using the more formal {\it k--means} statistical clustering as a benchmark classification.

\begin{table}[H]
\caption{Classification of announcements for the Eurozone volatility series using three alternative algorithms. The numbers in parentheses are the centers of the corresponding group. In the group {\it Plank} of the {\it SP--level} classification, the number in square brackets represents the cases with high volatility both at $t$ and $t-1$ ({\it High--Plank}).}
\begin{center}
\begin{tabular}{lccc|ccc}
Group&\multicolumn{1}{c}{{\it k--means}} &\multicolumn{1}{c}{{\it SP--level}}&\multicolumn{1}{c|}{{\it SP--diff}}&\multicolumn{1}{c}{{\it k--means}}&\multicolumn{1}{c}{{\it SP--level}}&\multicolumn{1}{c}{{\it SP--diff}}\\
\hline
&\multicolumn{3}{c|}{CAC40}&\multicolumn{3}{c}{DAX30}\\ [2mm]
Plank & 132      &  136    [0] & 136 & 136 & 140 [1] & 140    \\
      & (0.010)  &  (0.004)    &  (0.004) &(0.007) & (0.005) &(0.005) \\[2mm]
Squat & 5        &  1           &   1 & 3   &  0  & 0  \\
      & (-0.261) &  (-0.561)    & (-0.561) &(-0.198) & -  & - \\[2mm]
Jump  & 7        & 7            & 7 & 5       &   4& 4  \\
      & (0.704)  & (0.704)    &  (0.704) &(0.664) & (0.733)& (0.733) \\[2mm]
\hline

&\multicolumn{3}{c|}{FTSEMIB}&\multicolumn{3}{c}{IBEX35}\\ [2mm]
Plank & 118 & 125 [2] & 133& 123 & 131 [3] & 132\\
      &(0.010) & (0.018)& (0.035)&(0.020) & (0.008)& (0.005) \\[2mm]
Squat & 4& 2 & 1 & 9 & 2 & 1 \\
      & (-0.460) & (-0.606)& (-0.726) &(-0.309) & (-0.498) &(-0.526)  \\[2mm]
Jump & 22& 17 & 10 & 12& 11& 11  \\
     & (0.539) & (0.599)& (0.731) & (0.715) &(0.745) &(0.745)  \\[2mm]
\hline
\end{tabular}\label{tab:announ_regimechanges}
\end{center}
\end{table}

\begin{table}[H]
\caption{Evaluation of uncertainty by $U$ index for three alternative classification methods of four Eurozone volatility series.}
\center{
\begin{tabular}{lccc}
&{\it k--means}&{\it SP--level}&{\it SP--diff}\\ \hline
CAC40          & 0.124&0.088&0.088  \\[2mm]
DAX30 & 0.082&0.054&0.054  \\[2mm]
FTSEMIB  &   0.244&0.210&0.192\\[2mm]
IBEX35&0.207&0.156&0.154\\[2mm] \hline
\end{tabular} \label{tab:Uindex}
}
\end{table}

\begin{table}[H]
\caption{Adjusted Rand index between classification methods by volatility series.}
\center{
\begin{tabular}{lccc}
&{\it k--means/}&{\it k--means/}&{\it SP--level/}\\
&{\it SP--level}&{\it SP--diff}&{\it SP--diff}\\ \hline
CAC40          & 0.962&0.962&1  \\[2mm]
DAX30 & 0.962&0.962&1  \\[2mm]
FTSEMIB  &   0.925 &  0.851 &  0.919  \\[2mm]
IBEX35      &   0.922 &  0.913 &  0.990  \\[2mm]
\hline
\end{tabular} \label{tab:rand_methods}
}
\end{table}

\begin{table}[H]
\caption{Adjusted Rand index between pairs of volatility series by classification method.}
\center{
\begin{tabular}{lrrr}
&{\it k--means}&{\it SP--level}&{\it SP--diff}\\ \hline
CAC40/DAX30& 0.924&0.962&0.962  \\[2mm]
CAC40/FTSEMIB& 0.792&0.875&0.951  \\[2mm]
CAC40/IBEX35& 0.854&0.913&0.923  \\[2mm]
DAX30/FTSEMIB& 0.812&0.859&0.933  \\[2mm]
DAX30/IBEX35& 0.805&0.897&0.906  \\[2mm]
FTSEMIB/IBEX35& 0.741&0.828&0.912  \\[2mm]
\hline
\end{tabular} \label{tab:rand_countries}
}
\end{table}

\section{Concluding Remarks}
\label{sec:conclusion}
In this paper we derive a novel Markov Switching Multiplicative Error Model to include a component related to monetary policy actions: such a model extends the recent MEM--class contribution by \cite{Lacava:Gallo:Otranto:2020} which accommodates an additive component related to volatility dynamics induced by policy measures. As a relevant by--product, we advance a simple--to--obtain suggestion on how to map the information on estimated volatility regimes to classify announcements of a Central Bank in terms of their impact on volatility levels. Recent econometric literature is producing great efforts in evaluating these transmission mechanisms in real and financial economies, but generally they assume different announcements as producing the same effect.

In the model we propose within a more realistic scenario, the announcements are allowed to have different importance, and no prior classification is imposed. With  MS features, our model has the merit of extracting from volatility an unobservable signal attributable to the unconventional policy effects, with jumps in its intercept as a consequence of a policy  announcement: the estimated parameters allow us to derive a procedure to map the variations in the switching intercept into groups interpretable as policy effects on volatility.

We propose two smoothed probability--based classifications, one based on the thumb rule of the classification derived from the mode of the regime ({\it SP--level} method), and another based on the differences of the same smoothed probabilities (we call it {\it SP--diff}). The smoothed probabilities are a by--product of the Maximum Likelihood estimation, so they allow for an immediate classification of the announcements. The application on four European volatility series in the empirical application shows how the smoothed probability--based classifications provide a very similar clustering with respect to a statistical clustering  algorithm ({\it k--means}). Such a classification method based on the smoothed probabilities is extendable to all cases when a MS model is a suitable representation of the data.

Given the time series framework, the procedure could be extended to be adapted to the  real--time  classification of the announcements, when new observations are available, and use the framework in a forecasting context, not examined here. Given that the classification need to be reapplied to each new announcement, it remains to be seen how robust the previous classification is to the inclusion of new data or, for that matter, to outliers (see, for example, \cite{Durso:Degiovanni:Massari:2016}). Moreover, it could be interesting to verify whether such a classification method is robust with respect to alternative time--varying models, such as the smooth-transition MEM proposed in \cite{Gallo:Otranto:2015}, and alternative clustering methods.

{An interesting extension\footnote{We thank an anonymous referee to have suggested this possibility.} is to model the duration in a certain regimes as a function of the announcements. One way to investigate this concern is to adopt a time--varying transition probability (TVTP) matrix, where the probability to stay in the same regime at time $t$ depends on the state at $t-1$ and the dummy $\Lambda_t=1$ when there is an announcement at time $t$. As a consequence, we are able to estimate the duration of a certain state in the case of no announcement and in the case of announcement.
We did estimate such a TVTP MS--ACM and we indeed confirm that both probabilities, to stay in regime 0 and in regime 1, change in correspondence of the announcements, in the direction of a switch from regime 0 to 1 or the permanence in regime 1 (the coefficient  for $p_{11}$ is, however, not significant, in view perhaps of the short  duration of regime 1). This model outperforms the standard model in terms of Likelihood Ratio test, but, when used for classification purposes (the main goal  in this context), it introduces less sharp results relative to the MS--ACM case, as it provides smoothed probabilities often close to 0.5. To save space, these results are available in the supplemental material section.



 \bibliographystyle{elsarticle-num}
  \bibliography{biblio}





\end{document}